\documentclass[twocolumn,superscriptaddress,amsmath,amssymb,aps,prb]{revtex4}
\usepackage{graphicx}
\usepackage{dcolumn}
\usepackage{bm}
\usepackage{hyperref}
\usepackage{amsmath,amssymb}
\hypersetup{colorlinks=true, linkcolor=blue,	citecolor=blue,	filecolor=blue,	urlcolor=blue}

\begin{document}

\title{Evidence of electron correlation and weak bulk plasmon in SrMoO$_{3}$}

\author{Asif Ali}
\affiliation{Department of Physics, Indian Institute of Science Education and Research Bhopal, Bhopal Bypass Road, Bhauri, Bhopal 462066, India}%

\author{B. H. Reddy}
\altaffiliation[Present address: ]{Department of Physics, Government College (A), Rajahmundry 533105, India}
\affiliation{Department of Physics, Indian Institute of Science Education and Research Bhopal, Bhopal Bypass Road, Bhauri, Bhopal 462066, India}%

\author{Ravi Shankar Singh}
\email{rssingh@iiserb.ac.in}
\affiliation{Department of Physics, Indian Institute of Science Education and Research Bhopal, Bhopal Bypass Road, Bhauri, Bhopal 462066, India}%


\begin{abstract}
	We investigate the electronic structure of highly conducting perovskite SrMoO$_{3}$ using valence band photoemission spectroscopy and electronic structure calculations. Large intensity corresponding to coherent feature close to Fermi level is captured by density functional theory (DFT) calculation. An additional satellite  at $\sim$ 3 eV binding energy remains absent in DFT, hybrid functional (DFT-hybrid) and dynamical mean field theory (DFT + DMFT) calculations. Mo 4$d$ spectra obtained with different surface sensitive photoemission spectroscopy suggest different surface and bulk electronic structures. DFT + DMFT spectral function is in excellent agreement with the coherent feature in the bulk Mo 4$d$ spectra, revealing moderate electron correlation strength. A large plasmon satellite and signature of strong electron correlation are observed in the surface spectra, while the bulk spectra exhibits a \textit{weak} plasmon satellite.
\end{abstract}

\maketitle

	Transition metal oxides (TMOs) exhibit diverse physical phenomena such as  metal-insulator transition \cite{ImadaFujimori_MIT_RMP1998}, superconductivity \cite{Bednorz1986}, multiferroicity \cite{M.Dawber_Ferroelectricoxides_RevModPhys.77.1083}, giant and colossal magnetoresistance \cite{ M.B.Salamon_GMR_RevModPhys.2001}, non-Fermi liquid behaviour \cite{Pkhalifah, PKhalifah_CSRO_NFLPhysRevB2004}, quantum phase transition \cite{Z.Q.Mao_BaRu6O12_PRL2003} and various exotic magnetic orders \cite{PBAllen_SrRuO3_PhysRevB1996, RSPErry_MetamagnetismSr3Ru2O7_PRL2001, S.A.Grigera_Sr3Ru2O7_Science2004,C.-Q.Jin_FerromagnetismARuO3_PNAS2008, S.Trebst_Kitaev_PhysicsReports20221}. It is well recognized that electron correlation plays a crucial role in describing such exotic properties. The electron correlation is expected to be weak in 4$d$ TMOs compared to 3$d$ TMOs due to larger spatial extension of 4$d$ orbitals than that of 3$d$ orbitals. However, against this general belief, varying strength of electron correlation have been observed in 4$ d $ TMOs leading to exotic ground states \cite{G.Koster_SRO_RevModPhys2012, HDKim_Ru3d_PhysRevLett2004, S.Nakatsuji_CaSrRuO4_PhysRevLett.2000, Maeno1994, Y.Park_SrNbO3_CommPhys2020,B.Chiara_SrNbO_ARPES_PRM2020, KMaiti_CaSaRuO3_PRB2005,KarpJSr2MoO42020, YZhangSrRhO3PRB2020,M.W.Havekort_Sr2RhO4._PhysRevLett2008, BHReddy_APd3O4_JPCM2021}.

	Among 4$d$ TMOs, molybdenum based perovskites $A$MoO$_{3}$ ($A$ = Ca, Sr, Ba) exhibit metallic behaviour with Pauli paramagnetism \cite{HAYASHI_AMoO3_1979}. Exceptionally high conductivity is found in SrMoO$_{3}$ associated with highly delocalized Mo 4$d^{2}$ electrons with weak contribution from phonon scattering \cite{NagaiAPL2005}. Resistivity measurement shows a Fermi-liquid behaviour and the Sommerfeld coefficient evaluated from specific heat measurements is twice than that obtained from band structure calculation, suggesting an enhanced quasiparticle mass \cite{NagaiAPL2005}. Such quasiparticle mass enhancement is considered as signature of electron correlation, also observed in other strongly correlated systems \cite{Inoue_CSVO_PhysRevB1998, PBAllen_SrRuO3_PhysRevB1996}. SrMoO$_{3} $ exhibits two structural transitions with lowering temperature where the room temperature cubic $Pm\bar{3}m$ structure goes to tetragonal $I4/mcm$ structure and further to orthorhombic $Imma$ structure at around 266 K and 124 K, respectively \cite{Macquart2010}. The low temperature pseudo-cubic structures arise due to the rotation and/or tilting of MoO$_{6}$ octahedra. Phonon calculations within DFT + $U$ ($U$: on-site Coulomb repulsion) framework emphasizes that the electron correlation is responsible for the structural transitions \cite{J.-L.Hand_SrMoO3_PRM2021}. Also, DFT + DMFT correctly predicts the non-magnetic ground state along with octahedra rotation for the $ Imma $ structure \cite{AHampel_SrMoO3_PRB2021}. Quasiparticle spectral weight ($ Z $) is found to be $\sim$ 0.6 \cite{AHampel_SrMoO3_PRB2021, APaulPRB2019}, commensurate with specific heat measurements and recent angle-resolved photoemission spectroscopy (ARPES) measurement \cite{NagaiAPL2005, E.Cappelli_SrMoO3ARPES_PRM2022}. However, DFT + DMFT fails to capture an intense satellite observed at $\sim$ 2.5 eV binding energy in hard $ x $-ray valence band photoemission spectra. This satellite has been argued to have a plasmonic origin \cite{HWadatiSrMoO3}. Various GW calculations have also predicted a plasmon satellite at about 3 eV binding energy but with much smaller spectral weight in contrast to the experiment \cite{NilssonPRM2017, PetochhiPRR2020, TianyuCellDMFTPRX2021}.
 
	In this letter, we investigate the electronic structure of SrMoO$_{3}$ using valence band photoemission spectroscopy on \textit{in-situ} fractured polycrystalline sample. We observe considerably different surface and bulk electronic structure. Observation of significantly \textit{weak} plasmon satellite along with moderate electron correlation in the bulk Mo 4$d$ spectra, are consistent with many-body theoretical calculations. We further discuss the surface electronic structure.
 
	High quality polycrystalline sample of SrMoO${_3}$ was prepared by solid state reaction method using high purity MoO$_{3}$ (99.99 \%) and SrCO$_{3}$ (99.995 \%). Thoroughly ground mixture was palletized and heated at 600 $^{\circ}$C for 8 hours and further at 1250 $^{\circ}$C for 48 hours with intermittent grindings. Heat treatments were performed under 5$\%$ hydrogen mixed argon gas flow, resulting in very hard brick-red pellets. The phase purity and crystal structure were confirmed by \textit{x}-ray diffraction pattern collected at room temperature (Fig. S1 of supplemental material (SM) \cite{Supplemental}). The cubic lattice parameter was found to be 3.975(8) \AA ~in excellent agreement with earlier reports \cite{NagaiAPL2005,Macquart2010}. Photoemission spectroscopic measurements were carried out at 30 K using monochromatic Al $K_{\alpha}$ ($h\nu$ = 1486.6 eV) and He {\scriptsize II} ($h\nu$ = 40.8 eV) radiation (energy) on \textit{in-situ} fractured sample (base pressure $\sim$ 4$\times$10$^{-11}$ mbar). The Fermi level ($E_{F}$) and energy resolution were determined by measuring the Fermi cut-off of a clean polycrystalline silver at 30 K. The energy resolutions for Al $K_{\alpha}$ and He {\scriptsize II} spectra were set to $\sim$ 300 meV and $\sim$ 10 meV, respectively.

	Electronic structure calculations were performed for the orthorhombic \textit{Imma} structure with structural parameters adopted from Ref. \cite{Macquart2010}. Full potential linearized augmented plane wave method, as implemented in \textsc{wien2k} \cite{WIEN2k}, was used for the DFT calculation within generalized gradient approximation of Purdew-Burke-Ernzerhof \cite{PBEPRL1996}. For the DFT-hybrid calculations, screened hybrid functional (YS-PBE0) \cite{TranYSPBE0PRB2011} was constructed by replacing the $\alpha$ fraction of semi-local exchange with Hartree-Fock exchange, where $\alpha$ is the mixing parameter. The energy and charge convergence criteria were set to $10^{-5}$ eV and $10^{-4}$ electronic charge per formula unit (f.u.), respectively.
	Fully charge self-consistent DFT + DMFT calculation for 50 K ($\beta$ $\approx$ 232 eV$^{-1}$) were performed using eDMFT code with continuous time quantum Monte Carlo impurity solver and ``exact" double counting \cite{HauleDMFTPRB2010,Haule_CTQMC_PRB2007, HauleExactDCPRL2015}. A hybridization window of $\pm$ 10 eV was used and all five $d$ orbitals as basis was chosen for correlated Mo atoms. Analytical continuation was performed using maximum entropy method to calculate self-energy on the real axis \cite{HauleDMFTPRB2010}. The chosen local axes were nearly aligned with Mo-O bond direction of the MoO$_{6}$ octahedra. The Hubbard-$U$ and Hund's coupling $J$ were set to 6.0 eV and 0.7 eV, respectively. Total 4000 $k$-points were used for DFT and DFT + DMFT calculations and 500 $k$-points were used for DFT-hybrid calculation. Density of states (DOS) was calculated using 4000 $k$-points for all the calculations.

\begin{figure}[t]
	\centerline{\includegraphics[width=.45\textwidth]{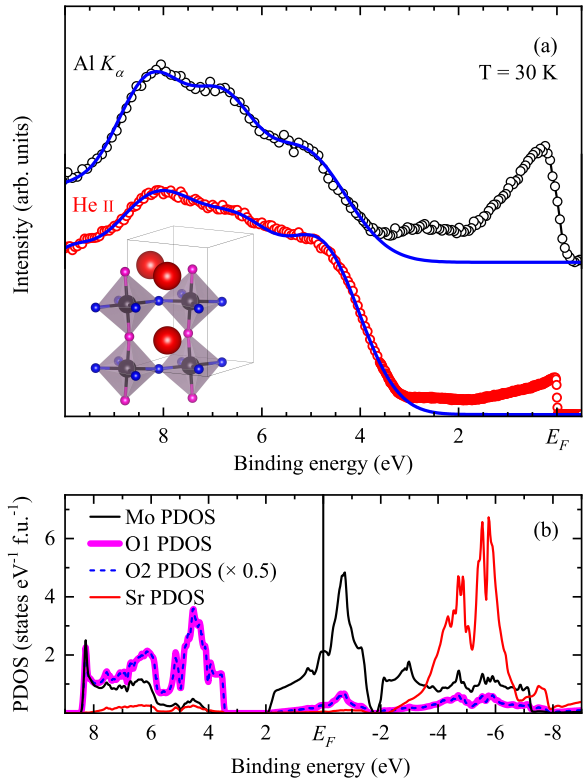}}
	\caption{(color online) (a) Valence band photoemission spectra of SrMoO$_{3}$ collected at 30 K using Al $K_{\alpha}$ (black circles) and He \textsc{ii} (red circles) radiations. The O 2$p$ band contributions (see text) are shown by blue line. Atomic arrangement in orthorhombic structure is shown where red, black, pink and blue spheres represents Sr, Mo, O1 (apical) and O2 (basal) atoms, respectively. (b) DFT calculated PDOS for Sr, Mo, O1 and O2 atoms for the orthorhombic $Imma$ structure.}\label{fig:Fig1}
\end{figure}

	Atomic arrangement in orthorhombic SrMoO$ _{3} $ has been shown in Fig \ref{fig:Fig1} (a). Each Mo atom is surrounded by two types of  crystallographically different oxygens (two apical (O1) and four basal (O2)), while the Sr atom sits in the void created by surrounding MoO$ _{6} $ octahedra. Valence band is formed by the hybridization between Mo 4$d$ and O 2$p$ states and contributions from Sr states are expected to be negligible in the occupied energy range. Valence band photoemission spectra of SrMoO$_{3}$ collected at 30 K using Al $K_{\alpha}$ and He \textsc{ii} radiations have been shown in Fig. \ref{fig:Fig1} (a). Both spectra exhibit two distinctly separated groups of features below and above 3.5 eV binding energy. Considering the larger photo-ionization cross-section ratio of Mo 4$d$ states with O 2$p$ states in case of Al $K_{\alpha}$ than in case of He \textsc{ii} \cite{YEH19851}, the features below 3.5 eV binding energy can be attributed to Mo 4$d$ states and the features above 3.5 eV binding energy can be attributed to O 2$p$ states. 

	The attributed characters of the spectral features are further examined using DFT calculation. Calculated partial density of states (PDOS) for Sr, Mo, O1 and O2 atoms are shown in Fig. \ref{fig:Fig1} (b). It is clear that the valence band is formed by Mo 4$d$ and O 2$p$ hybridized states, while Sr states have negligible contribution in the occupied region (appearing between -2 eV to -8 eV). There are three sets of features above $\sim$ 3.5 eV binding energy centered $\sim$ 4.5 eV corresponding to non-bonding states primarily having O 2$ p $ character and $\sim$ 6 eV and $\sim$ 8 eV features corresponding to bonding states with Mo 4$ d $ and O 2$ p $ mixed character. The anti-bonding states located between 2 eV to -2 eV binding energy primarily have Mo 4$d$ $t_{2g}$ character, while the Mo 4$d$ $e_{g}$ states appear between -2 eV to -7.2 eV binding energy. The average $\angle$ Mo-O-Mo reduces to $\sim$ 172.6$^{\circ}$ ($\angle$Mo-O1-Mo = $\sim$ 171.3$^{\circ}$ and $\angle$Mo-O2-Mo = $\sim$ 173.2$^{\circ}$) in the orthorhombic $Imma$ structure from 180$^{\circ}$ in the cubic $ Pm\bar{3}m$ structure \cite{Macquart2010}. It is to note here that the PDOS corresponding to O1 and O2 ($\times$ 0.5) are almost degenerate, as evident from the figure, suggesting negligible influence of the octahedra rotation and/or tilt in the electronic structure of SrMoO$_{3}$. This is also confirmed by very similar calculated DOS and essentially similar valence band spectra collected at 300 K and 30 K using Al $K_{\alpha}$, (Fig. S2 of SM \cite{Supplemental}). 

\begin{figure}[t]
	\centerline{\includegraphics[width=.45\textwidth]{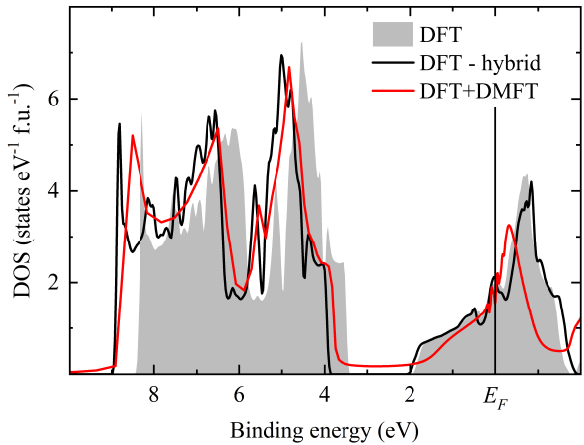}}
	\caption{(color online) Total DOS of orthorhombic SrMoO$_{3}$ calculated using DFT, DFT-hybrid and DFT + DMFT.}\label{fig:Fig2}
\end{figure}

	The overall comparison of experimental spectra with DFT results reveal two significant differences, (i) a mismatch in the energy position of the O 2$p$ band and (ii) an additional satellite feature around 3 eV binding energy in the experimental spectra. The feature at $\sim$ 5 eV binding energy in the experimental spectra appears at about 4.5 eV in the DOS calculated within DFT. This overestimation of O 2$p$ band energies has been attributed to the underestimation of electron correlation in DFT, as also observed in various TMOs \cite{SarmaDD_PRL1995}.
	The methods beyond DFT, such as DFT-hybrid and DFT + DMFT, have been found to be quite successful in case of TMOs with varying strength of electron correlation. The hybrid functionals containing some part of the Hartree-Fock exchange have been very successful in correctly describing the electronic properties of many semiconductors \cite{TranYSPBE0PRB2011,HSE2004, BHReddy_APd3O4_JPCM2021}, insulators \cite{TranYSPBE0PRB2011, JPaier_hybriddielectric_PRBR2008, SBansal_Li2RuO3_JPCM2021} and correlated metallic systems \cite{MCasedei_Cerium_PRB2016, Safdat_PbMoO3_2017, FalkePRB2021}. The mixing parameter $\alpha$ is related to the dielectric properties of a material and have been varied in order to match with the experimental results \cite{Koller_JPCM2013} even in case of metals \cite{FalkePRB2021}. Indeed, a shift of about 0.5 eV towards higher binding energy in the O 2$p$ band position is found for $\alpha$ = 0.10 as shown in the Fig. \ref{fig:Fig2}. DFT-hybrid calculations for cubic structure with varying $\alpha$ from 0.25 to 0.10 has been shown in SM \cite{Supplemental}. To further examine the observed shift, we have also performed DFT + DMFT calculation. The DFT + DMFT has emerged as a successful method to treat weak to strong electron correlation in many $d$ and $f$ electron systems \cite{GKotliar_RevModPhys.78.865, G.Kotliar_PhysicsToday2004}. The total DOS calculated using DFT + DMFT has been shown in Fig. \ref{fig:Fig2} which also exhibits shift of the O 2$p$ band towards higher binding energy as compared to DFT, consistent with the experimental spectra and DFT-hybrid. Reduced Mo 4$d$ bandwidth in DFT + DMFT suggests strong renormalization with quasiparticle weight $Z$ $\approx$ 0.5 (for $t_{2g}$ orbitals), consistent with earlier calculations and specific heat measurements \cite{APaulPRB2019, AHampel_SrMoO3_PRB2021, E.Cappelli_SrMoO3ARPES_PRM2022, NagaiAPL2005}. The DFT + DMFT provides a better description of the experimental spectra, as discussed later.

\begin{figure}[t]
	\centerline{\includegraphics[width=.45\textwidth]{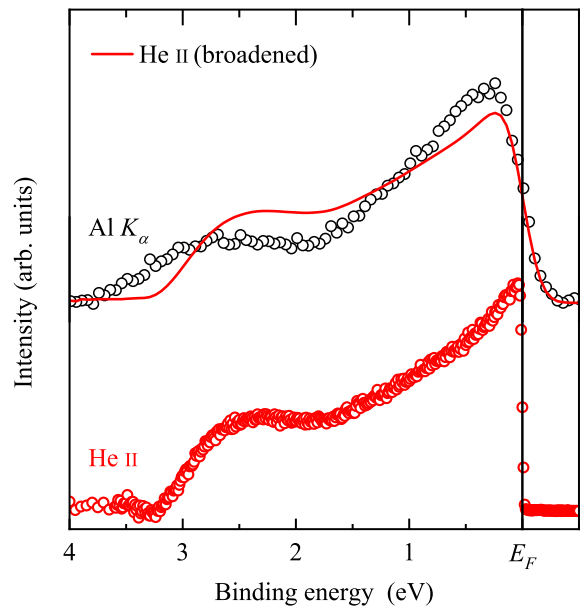}}
	\caption{(color online) Extracted Mo 4$d$ band of SrMoO$ _{3} $ from Al $K_{\alpha}$ (black circles) and He \textsc{ii} (red circles) valence band spectra. The resolution broadened He \textsc{ii} spectra is shown by red line.}\label{fig:Fig3}
\end{figure}

	Interestingly, all of the calculations discussed above fails to describe the $\sim$ 3 eV satellite feature of the experimental spectra. Such higher binding energy feature observed in the photoemission spectra is a typical signature of correlation induced lower Hubbard band (LHB) in strongly correlated TMOs and has been successfully captured by DFT + DMFT \cite{A.Sekiyama_CaSrVO_PRL2004,Silke_VO2cDMFT_PRL2005, Z.V.Pchelkina_LRODMFT_PRB2015}. However, in the present case the satellite feature can not be attributed to LHB since it remains absent in the DFT + DMFT. A similar feature observed in earlier photoemission experiment was suggested to be a plasmon satellite \cite{HWadatiSrMoO3}, as also observed in subsequent GW calculations \cite{NilssonPRM2017, PetochhiPRR2020,TianyuCellDMFTPRX2021}.

\begin{figure}[t]
	\centerline{\includegraphics[width=.45\textwidth]{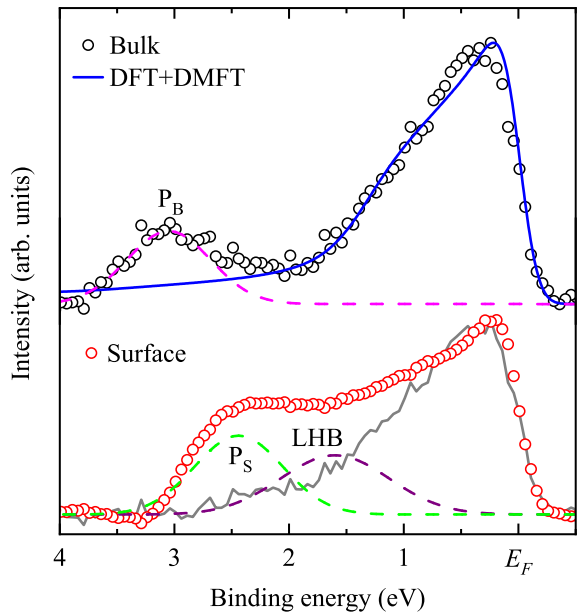}}
	\caption{(color online) Upper panel shows bulk spectra (open circles) of SrMoO$_{3}$. The solid blue and dashed pink lines shows the DFT + DMFT spectral function and Gaussian type bulk plasmon satellite (P$ _{B} $), respectively. Lower panel shows surface spectra (open circles) of SrMoO$ _{3} $. The solid grey, dashed green and dashed maroon lines shows bulk coherent feature, surface plasmon satellite (P$ _{S} $) and surface LHB, respectively.}\label{fig:fig4}
\end{figure}

	In order to further investigate, we extract the Mo 4$d$ band contributions from the valence band spectra. This can be reliably done by subtracting the O 2$p$ band features (simulated using three Gaussian and shown as blue lines), since they are distinctly separate from Mo 4$d$ band in experimental and calculated valence band shown in Fig. \ref{fig:Fig1}. The extracted Mo 4$d$ bands, normalized by total integrated intensity, are shown in Fig. \ref{fig:Fig3}, for both Al $K_{\alpha}$ and He \textsc{ii} spectra. Both spectra exhibit intense feature close to $E_{F}$ with a hump like satellite feature. The relatively weak satellite feature observed here is consistent with our previous $x$-ray photoemission study on \textit{ex-situ} thin film \cite{AAli_SMOfilm_AIPConf2019}. He \textsc{ii} spectra exhibits a sharper Fermi cut-off due to higher resolution than that of Al $K_{\alpha}$ spectra. For a direct comparison, we broaden the He \textsc{ii} spectra by a Gaussian of 0.3 eV width ($\sim$ energy resolution of Al $K_{\alpha} $ spectra) as shown by line. Now, these two spectra having similar resolution broadening differ from each other only in probing depth employed in the photoemission spectroscopy where Al $K_{\alpha}$ spectra is more bulk sensitive while He \textsc{ii} spectra is more surface sensitive. Different line shape of these two spectra suggests that the surface and bulk electronic structures are distinctly different in this system. Lowered symmetry, reduced coordination number and/or surface reconstruction $ etc.$ at the surface may lead to difference in the electronic structure. Thus, it is essential to disentangle the contribution of the surface to understand the intrinsic bulk electronic structure \cite{KMaiti_LACaVO_PRL1998,KMaiti_CaSaRuO3_PRB2005}.

	Photoemission spectral intensity for the incident photon energy $h\nu$ can be expressed as $I_{h\nu}(E) = e^{-l/\lambda_{h\nu}}f_{b}(E)+ (1-e^{-l/\lambda_{h\nu}})f_{s}(E)$, where $l$ is surface layer thickness, $\lambda$ is photoelectron mean free path (probing depth) and $f_{b}(E)$ and $f_{s}(E)$ represent bulk and surface spectra, respectively. $f_{b}(E)$ and $f_{s}(E)$ were estimated using $l/\lambda_{Al K_{\alpha}}$ = 0.45 and $l/\lambda_{He \textsc{ii}}$ = 1.75 and have been shown in Fig. \ref{fig:fig4}. The obtained spectra are quite robust with respect to the energy positions and relative intensity of features within 20$\%$ variation of $l/\lambda$, providing confidence in the analysis \cite{KMaiti_LACaVO_PRL1998,KMaiti_CaSaRuO3_PRB2005}.

	As evident from Fig. \ref{fig:fig4}, surface and bulk spectra (normalized by total integrated intensity) are distinctly different. The intrinsic bulk spectra exhibiting an intense coherent feature below 2 eV binding energy is compared with resolution broadened occupied part of the DFT + DMFT spectral function (matched at highest intensity). An excellent agreement for the coherent feature confirms moderate electron correlation in this system. A significantly \textit{weak} satellite (P$ _{B} $) in the bulk spectra at $\sim$ 3 eV binding energy is commensurate with the broad and weak plasmon satellite observed in various GW calculations \cite{NilssonPRM2017, PetochhiPRR2020, TianyuCellDMFTPRX2021}. It is to note here that the DFT, DFT-hybrid and DFT + DMFT calculations, performed here, fail to describe the plasmonic satellite, since, these calculations do not incorporate any long-range (non-local) interactions which are essential for description of plasmonic excitations \cite{HWadatiSrMoO3,Boehnke_PRB2016}.

	The surface spectra exhibits significantly reduced total width and enhanced satellite feature (appearing below 2.5 eV) in contrast to the bulk spectra. Interestingly, the surface spectra obtained here is strikingly similar to hard $x$-ray photoemission spectra on \textit{ex-situ} thin films \cite{HWadatiSrMoO3}. Plasmon satellites in the core level photoemission have been extensively studied revealing different lineshape and intensity for the bulk and surface plasmon. Surface plasmon satellite is expected to appear at smaller energy ( $\sqrt{2}$ times lower) than bulk plasmon satellite due to reduced dimensionality and/or confinement effects \cite{R.H.Ritchie_PhysRev.106.874_1956, Inglesfield_1983}. These have indeed been observed in grazing angle core-level photoemission experiments \cite{Biswas_AlPlasmon_PRB2003} while the enhancement of surface plasmon satellites are attributed to the enhancement of extrinsic plasmon losses at the surface \cite{Yubero_calculationAlplasmon_PRB2005, YUBERO_Siplasmon_SUrfScience2005}. A careful look at the surface spectra reveals that the coherent feature remains at very similar position while an intense and shifted satellite suggests an enhanced surface plasmon (P$ _{S} $), roughly $\sqrt{2}$ times lower energy than the bulk plasmon (P$ _{B} $). Deconvolution of the surface spectra as shown in Fig. \ref{fig:fig4}, requires at least two Gaussian type features in addition to bulk coherent feature (obtained from bulk spectra after subtracting Gaussian type P$ _{B} $ and matched at highest intensity), suggesting an additional feature at $\sim$ 1.5 eV binding energy (Gaussian peaks have been used for simplistic illustration). This additional feature can be a signature of LHB appearing due to enhanced effective electron correlation at the surface and requires further investigations on a high-quality crystals/thin films.

	In conclusion, we have investigated the electronic structure of SrMoO$_{3}$ using valence band photoemission spectroscopy. An accurate description of the O 2$p$ band position is found with DFT-hybrid and DFT + DMFT calculations. Valence band obtained using different photon sources suggest that the surface and bulk electronic structures are quite different in this system. The large coherent feature in bulk spectra is commensurate with DFT + DMFT spectral function suggesting moderate electron correlation in the Mo 4$d$ orbitals. Enhanced plasmon satellite and signature of strong correlation induced LHB are observed in the surface spectra. Intrinsic bulk spectra exhibits a \textit{weak} plasmon satellite as also observed in various GW calculations.

	We thank S. K. Pandey (IIT Mandi) for fruitful discussions. We acknowledge the support of CIF and HPC facilities at IISER Bhopal. We also thankfully acknowledge the funding from DST-FIST (Project No. SR/FST/PSI-195/2014C).

%

\end{document}


\title{Supplemental Material for ``Evidence of electron correlation and weak bulk plasmon in SrMoO$_{3}$"
	}
	
\author{Asif Ali}
\affiliation{Department of Physics, Indian Institute of Science Education and Research Bhopal, Bhopal Bypass Road, Bhauri, Bhopal 462066, India}%

\author{B. H. Reddy}
\altaffiliation[Presently at ]{Department of Physics, Government College (A), Rajahmundry 533105, India}
\affiliation{Department of Physics, Indian Institute of Science Education and Research Bhopal, Bhopal Bypass Road, Bhauri, Bhopal 462066, India}%

\author{Ravi Shankar Singh}
\email{rssingh@iiserb.ac.in}
\affiliation{Department of Physics, Indian Institute of Science Education and Research Bhopal, Bhopal Bypass Road, Bhauri, Bhopal 462066, India}%
		
	\maketitle	

\begin{figure}[H]
	\centering
	\includegraphics[width=.4\textwidth]{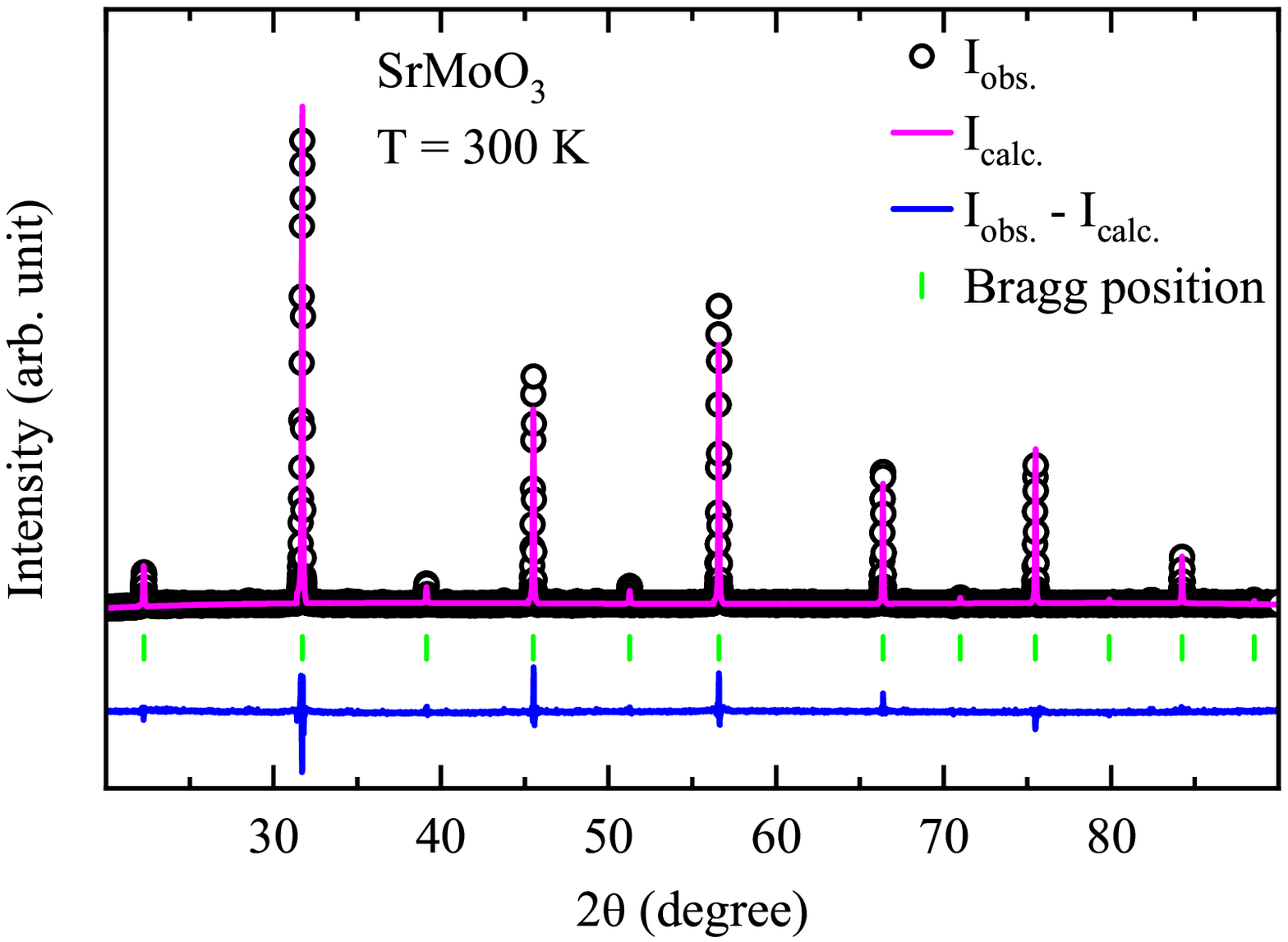}
	\caption{\textbf{Crystal structure:} Room temperature powder $x$-ray diffraction pattern (I$_{\text{obs.}}$) of SrMoO$_{3}$ collected using Cu $K_{\alpha}$ radiation ($\lambda$ = 1.5406 \AA) along with Rietveld refinement result (I$_{\text{calc.}}$) with $Pm\bar{3}m$ space group. The blue line and green vertical bars represents difference (I$_{\text{obs.}}$ - I$_{\text{calc.}}$) and Bragg reflection position, respectively.}\label{fig:XRD}
\end{figure}

\begin{figure}[H]
	\centering
	\includegraphics[width=.77\textwidth]{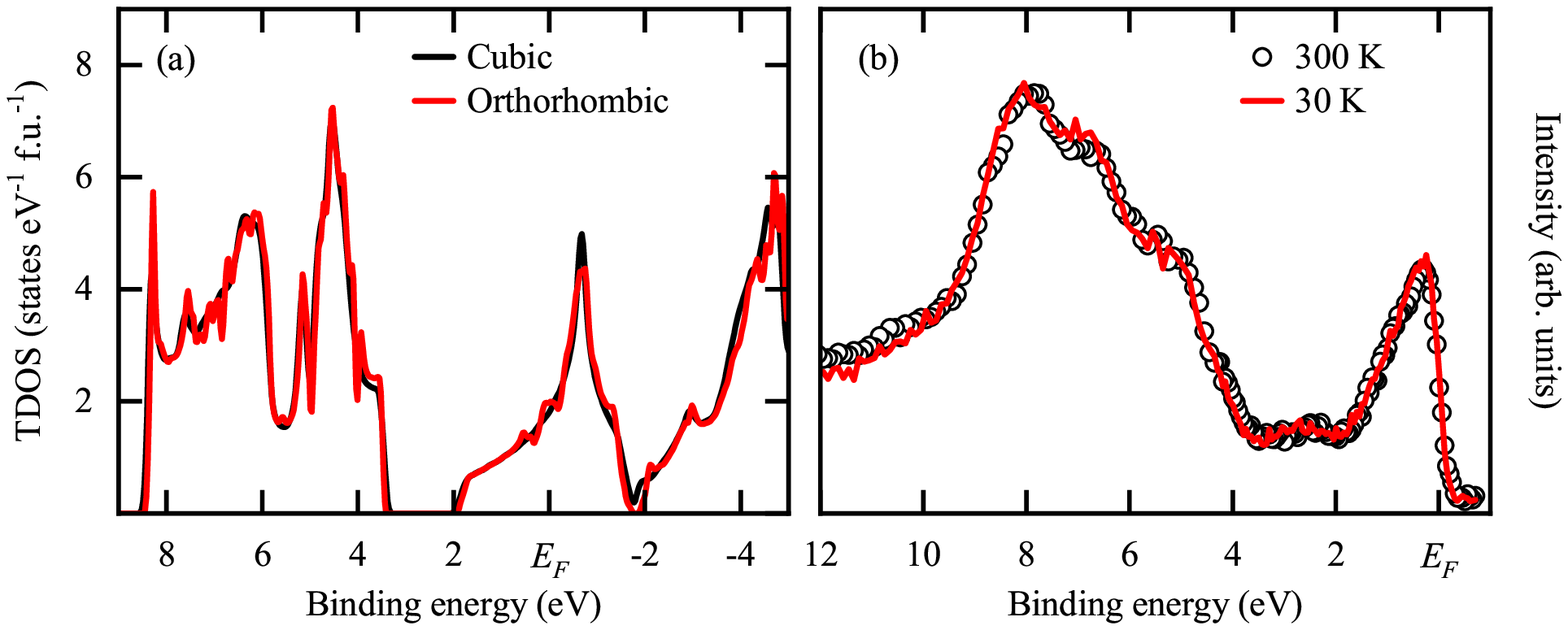}
	\caption{\textbf{DFT and Al $ K_{\alpha} $ spectra:} (a) DFT calculated total density of states (TDOS) for cubic and orthorhombic structure of SrMoO$ _{3}$. (b) Valence band photoemission spectra collected at 300 K (circles) and 30 K (line) using Al $ K_{\alpha} $ radiation. Essentially similar TDOS and experimental spectra for cubic and orthorhombic  structures indicated very similar electronic structures.} \label{fig:SFig2}
\end{figure}

\begin{figure}[H]
	\centering
	\includegraphics[width=.65\textwidth]{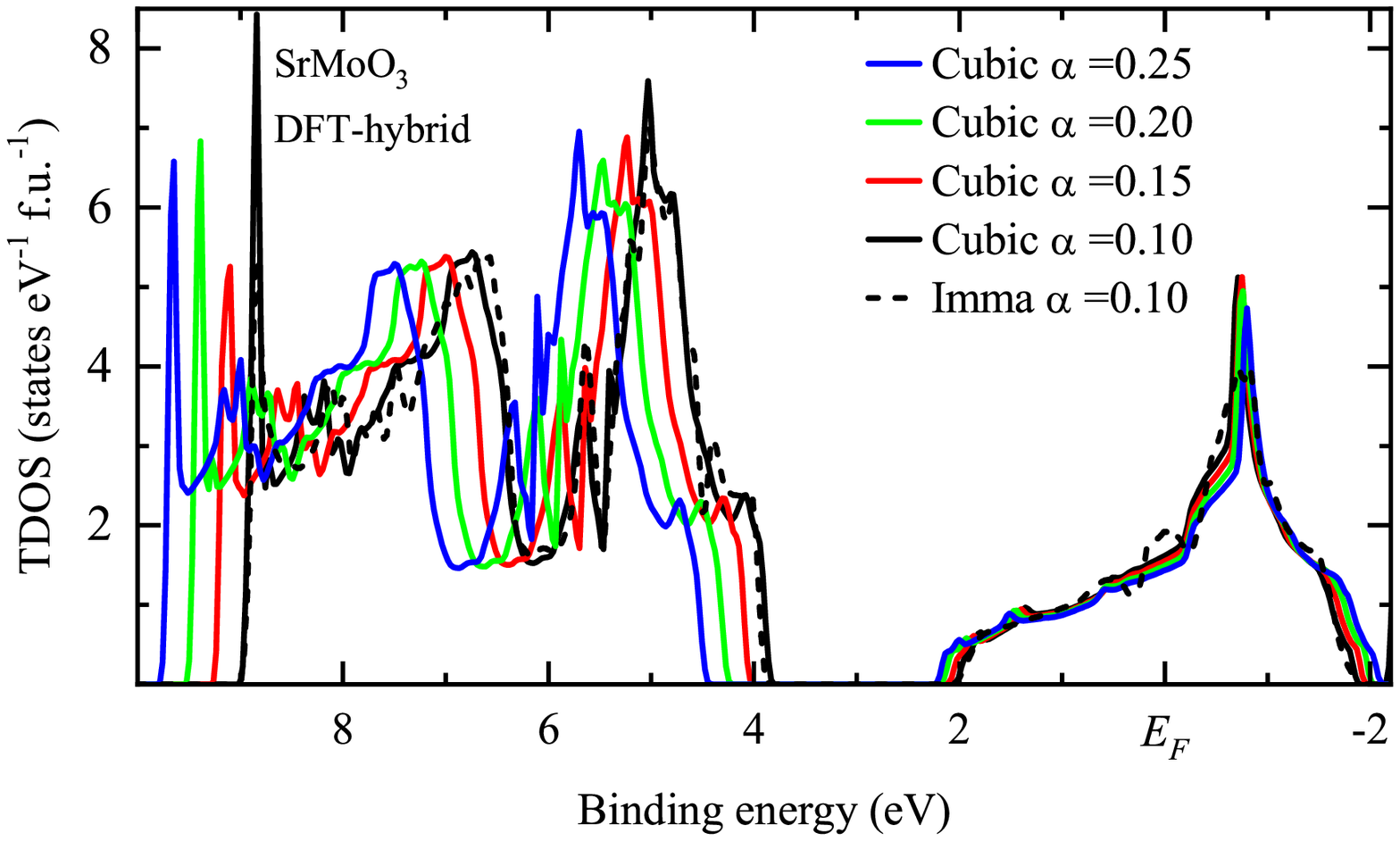}
	\caption{\textbf{DFT-hybrid:} DFT-hybrid calculated TDOS for cubic structure with different $\alpha$ (solid lines) and for orthorhombic ($Imma$) structure with $\alpha$ = 0.10 (dashed line). \\	
	Since the electronic structure for the cubic and orthorhombic structures remain very similar, we use the cubic structure for computationally expensive DFT-hybrid calculations with varying $\alpha$. The equivalent cubic lattice parameter $a$ = (volume of orthorhombic cell/4)$^{1/3}$ = 3.969 \AA, was used for the calculations with total 1000 $k$ points. The results are shown by solid lines. The O 2$p$ band starts from $\sim$ 4.5 eV binding energy for $\alpha$ = 0.25 and moves towards lower binding energy with decreasing $\alpha$. The experimental O 2$p$ band shown in Fig. 1 of main paper, is well reproduced for $\alpha$ = 0.10 -0.15. The results for the cubic and orthorhombic structure are very similar as confirmed with calculations for orthorhombic structure for $\alpha$ = 0.10, which have been presented in the Fig. 2 of main paper.} \label{fig:SFig3}
\end{figure}